\documentclass[10pt,aps,prl,twocolumn,superscriptaddress,preprintnumbers]{revtex4-1}
\usepackage{tabularx} 
\usepackage{graphicx} 
\usepackage{hyperref}  
\usepackage{amssymb}  
\usepackage{amsmath}  
\usepackage[]{units}
\usepackage[utf8]{inputenc}
\begin{document}
\title{Precise determination of proton magnetic radius from electron scattering data}
\preprint{JLAB-THY-20-3149}
\author{J.~M.~Alarcón}
\affiliation{Departamento de Física Teórica \& IPARCOS, 
Universidad Complutense de Madrid, 28040 Madrid, Spain}
\author{D.~W.~Higinbotham}
\affiliation{Jefferson Lab, Newport News, VA 23606}
\author{C.~Weiss}
\affiliation{Jefferson Lab, Newport News, VA 23606}
\begin{abstract}
We extract the proton magnetic radius from the high-precision electron-proton elastic scattering 
cross section data. Our theoretical framework combines dispersion analysis and chiral 
effective field theory and implements the dynamics governing the shape of the low-$Q^2$ form factors. 
It allows us to use data up to $Q^2\sim$ 0.5 GeV$^2$ for constraining the radii and overcomes 
the difficulties of empirical fits and $Q^2 \rightarrow 0$ extrapolation. We obtain a magnetic radius 
$r_M^p$ = 0.850 $\pm$0.001 (fit 68\%) $\pm$0.010 (theory full range) fm, 
significantly different from earlier results obtained from the same data,
and close to the extracted electric radius $r_E^p$ = 0.842 $\pm$0.002 (fit) $\pm$0.010 (theory) fm.
\end{abstract}
\maketitle
\section{Introduction}
The electromagnetic form factors (EM FFs) are the most basic expressions of the nucleon's 
finite spatial extent and composite internal structure. They describe the elastic response 
to external electric and magnetic fields as a function of the 4-momentum transfer $Q^2$ 
and can be associated with the spatial distributions of charge and current in nucleon.
The traditional representation of FFs in terms of 3-dimensional spatial 
densities at fixed instant time $x^0 =$ const.\ is appropriate only for nonrelativistic systems 
such as nuclei \cite{Miller:2018ybm}. For relativistic systems such as hadrons, the spatial 
structure is expressed through 2-dimensional transverse densities at fixed light-front time 
$x^+ = x^0 + x^3 =$ const. In the context of QCD these transverse densities can be regarded
as projections of the nucleon's partonic structure (generalized parton distributions)
\cite{Diehl:2003ny,Belitsky:2005qn,Boffi:2007yc}. The EM FFs thus
reveal aspects of the spatial distribution of quarks and their orbital motion and spin and 
have become objects of great interest in nucleon structure studies \cite{Punjabi:2015bba,Pacetti:2015iqa}.

The value of the electric and magnetic proton FFs at $Q^2 = 0$ is given by the total charge and magnetic 
moment of the proton, $G_E^p(0) = 1, G_M^p(0) = \mu^p = 2.793$.
The leading information about the spatial structure is in the first derivatives of the FFs at 
$Q^2 = 0$. They are conventionally expressed in terms of the equivalent electric and magnetic 
3-dimensional root-mean-square radii,
\begin{align}
\frac{d G_E^p}{dQ^2}(0) \; = \; -\frac{(r_E^p)^2}{6}, 
\hspace{2em}
\frac{1}{\mu^{p}} \frac{d G_M^p}{dQ^2}(0) \; = \; -\frac{(r_M^p)^2}{6};
\label{r2}
\end{align}
this does not imply an actual physical interpretation in terms of 3-dimensional densities;
the proper interpretation in terms of 2-dimensional densities is discussed below \cite{Miller:2018ybm}.
Besides their importance for nucleon structure, the FF derivatives are needed in tests of atomic 
bound-state calculations in quantum electrodynamics and in precision measurements of the Rydberg constant
\cite{Eides:2000xc,Pohl:2010zza}.

The proton electric (or charge) radius is extracted from the proton FFs measured in electron-proton 
elastic scattering, and from the nuclear corrections to atomic energy levels (electronic and muonic hydrogen)
measured in precision spectroscopy experiments; see Refs.~\cite{Pohl:2013yb,Carlson:2015jba} for a review.
Apparent discrepancies between the different extraction methods (``proton radius puzzle'') have
engendered intense experimental and theoretical efforts, including dedicated new FF measurements 
at low $Q^2$ using electron and muon beams \cite{Xiong:2019umf,Gilman:2017hdr}. 
Recent results seem to converge around 
$r_E^p = 0.84$ fm \cite{Higinbotham:2015rja,CODATA2018,Xiong:2019umf,Bezginov:2019mdi}.
The magnetic radius can be extracted only from elastic scattering measurements. 
Recent determinations based on the Mainz A1 data \cite{Bernauer:2010wm,Bernauer:2013tpr}, 
using methods developed in the context of the charge radius extraction, have resulted in a range of 
values that disagree with each other, $r_M^p =$ 0.78(2) fm \cite{Bernauer:2010wm}, 
0.914(35)fm and 0.776(38)fm \cite{Lee:2015jqa}, and significantly depart from older results 
$\sim$0.85 fm \cite{Belushkin:2006qa}. It is necessary to resolve these discrepancies and determine
the proton magnetic radius with an overall accuracy and consistency commensurate with those achieved
in the electric radius.

Here we report an extraction of the proton magnetic radius from electron scattering data using 
a novel theoretical framework based on dispersion analysis and chiral effective field theory
(DI$\chi$EFT) \cite{Alarcon:2018zbz,Alarcon:2017ivh,Alarcon:2017lhg,Alarcon:2018irp}.
It implements analyticity and the dynamics governing the shape of the low-$Q^2$ FFs
and allows us to use data up to $Q^2\sim$ 0.5 GeV$^2$ for constraining the radii,
increasing the sensitivity to the magnetic FF. It overcomes the difficulties 
in extraction methods based on empirical fits and $Q^2 \rightarrow 0$ 
extrapolation (functional form bias, unstable extrapolation),
particularly the issues related to the normalization of data sets taken at different 
incident energies. DI$\chi$EFT was used in Ref.\cite{Alarcon:2018zbz} to extract $r_E^p$
from an empirical FF parameterization \cite{Ye:2017gyb} and delivered a value of 
$r_E^p$ = 0.844(7) fm, as accepted in the CODATA 2018 update and confirmed 
by more recent measurements \cite{CODATA2018,Xiong:2019umf,Bezginov:2019mdi}.
In this work we use the method to extract both $r_M^p$ and $r_E^p$ from a direct analysis
of the cross section data, dominated by the Mainz A1 data \cite{Bernauer:2010wm,Bernauer:2013tpr}.
We obtain $r_M^p$ = 0.850 $\pm$0.001 (fit 68\%) $\pm$0.01 (theory full range) fm,
significantly different from the values extracted 
from the same data using other methods \cite{Bernauer:2010wm,Lee:2015jqa}, 
and surprisingly close to $r_E^p$. In the course we also improve our extraction of $r_E^p$
and verify the robustness of the results.
\section{Method}
DI$\chi$EFT is a method for calculating nucleon FFs combining dispersion analysis and chiral 
effective field theory. The theoretical foundations are described 
in detail in Refs.~\cite{Alarcon:2017ivh,Alarcon:2017lhg,Alarcon:2018irp}; 
applications to FF fits are discussed in Ref.~\cite{Alarcon:2018zbz}.
The FFs are represented as dispersion integrals over $t \equiv - Q^2$. The spectral functions 
on the two-pion cut at $t > 4 M_\pi^2$ are calculated using (i)~the elastic unitarity relation;
(ii)~$\pi N$ amplitudes computed in $\chi$EFT at leading order, next-to-leading order, and
partial next-to-next-to-leading order accuracy; (iii)~the timelike pion FF measured in $e^+e^-$ 
annihilation experiments. The approach includes $\pi\pi$ rescattering effects
and the $\rho$ resonance and generates accurate spectral functions up to 
$t \sim 1$ GeV$^2$. Higher-mass $t$-channel states are described by effective poles.
The parameters specifying the dynamical input (the low-energy constants of the $\chi$EFT calculation, 
and the strength of the effective poles) are related by the sum rules of dispersion theory 
and can be expressed in terms of the nucleon charges, magnetic moments, and radii.
For each (assumed) value of $r_E^p$ and $r_M^p$ the theory thus generates a unique prediction
for $G_E^p(Q^2)$ and $G_M^p(Q^2)$ with controlled theoretical uncertainties; see
Ref.~\cite{Alarcon:2018zbz} for a summary plot. It predicts the 
``shape'' of the spacelike FF as determined by analyticity (position of singularities) and 
dynamics (strength of singularities). In this way the values of the radii are correlated with the
predicted behavior of the FFs at finite momentum transfers $Q^2 \sim$ 1 GeV$^2$, 
allowing the use of such data for radius extraction. A computer code generating the DI$\chi$EFT
FF predictions and further information are available \footnote{See Supplementary Materials at [URL] 
for a Jupyter Notebook generating the DI$\chi$EFT FF predictions \cite{Alarcon:2018zbz}
and preparing reference plots.}.

%
% FIGURE
%
\begin{figure}[t]
\includegraphics[width=.75\columnwidth]{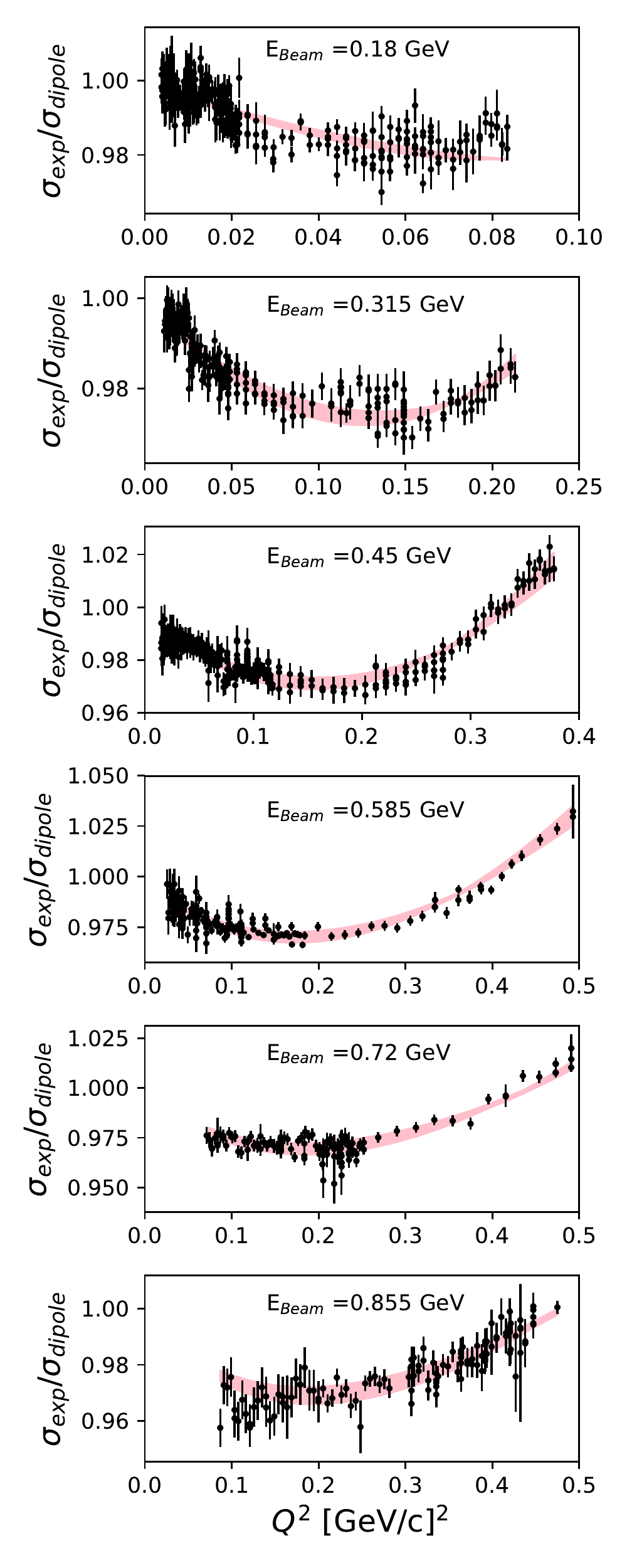}
\caption{Data: Mainz A1 electron-proton elastic scattering cross section data \cite{Bernauer:2010wm,Bernauer:2013tpr},
with the normalization of sets determined by our fit.
Band: Theoretical model (DI$\chi$EFT) with parameters $(r_E^p, r_M^p)$ obtained from our best fit.
The band shows the range of the model predictions obtained by varying the parameters in the 68\% 
confidence interval of the fit; it does not include the intrinsic theoretical uncertainty 
of the model \cite{Alarcon:2018zbz}.
Both data and model are divided by the cross section evaluated with the dipole FFs
($\Lambda^2 =$ 0.71 GeV$^2$).}
\label{fig:data}
\end{figure}

For our radius extraction we use the high-precision data in electron-proton elastic scattering
from the Mainz A1 experiment, which dominate the world data \cite{Bernauer:2010wm,Bernauer:2013tpr}. 
The experiment measured the elastic scattering cross section at momentum transfers 
$0.003 \lesssim Q^2 \lesssim 1$ GeV$^2$ and incident electron energies $E_{\rm Beam}$ 
from 0.18 to 0.855 GeV. The 2-dimensional data cover the cross section 
at a fixed $Q^2$ at various values of the virtual photon polarization parameter, $\epsilon$,
and allow for separation of the contributions of $G_E^p(Q^2)$ and $G_M^p(Q^2)$ through global fits,
generalizing the traditional Rosenbluth method \cite{Bernauer:2010wm,Bernauer:2013tpr}.

An important issue in global fits is the normalization of the data sets taken at different energies.
The normalization of the cross section data (both absolute and relative, between different energies)
is limited by the knowledge of the absolute luminosity in the different settings and subject to
considerable uncertainties. The combination of data taken at different energies therefore requires 
rescaling of the data sets, which depends on the functional form of the FFs or on other assumptions. 
In the context of empirical fits the effect of the rescaling on the random uncertainties of the data 
was studied in detail in Refs.~\cite{Bernauer:2013tpr,Lee:2015jqa}.
In the context of our approach this problem is naturally solved by the fact that the theoretical model 
predicts the shape of the FFs at finite $Q^2$ (in dependence of the radii). We can therefore perform a 
global fit with floating normalizations of the data sets, which can adjust themselves to the theoretical model; 
the physical information is in the variation of the data with $Q^2$, which tests the theoretical predictions 
for the shape and fixes the radius parameters through the best fit.

As the figure-of-merit for the global fit with floating normalizations we use a $\chi^2$ function
of the form
\begin{eqnarray}
\chi^2 &\equiv&  \chi^2 (r_E^p, r_M^p; \Lambda_1, ..., \Lambda_{N_{\rm set}})
\nonumber
\\[1ex]
&\equiv& {N_{\rm dat}^{-1}}
\sum_{\textrm{data $i$}} \left[ \frac{ \sigma_{{\rm thy}, i} - \Lambda_{k(i)} \, \sigma_{{\rm exp}, i}}
{\Lambda_{k(i)} \, \Delta\sigma_{{\rm exp}, i}} \right]^2 ,
\label{chi2}
\\[1ex]
\sigma_{{\rm thy}, i} &\equiv& \sigma(E_i, Q^2_i) [\textrm{DI$\chi$EFT, params $r_E^p, r_M^p$}].
\end{eqnarray}
The summation is over the $N_{\rm dat}$ data points labeled by $i$. $\sigma_{{\rm thy}, i}$ is the theoretical 
electron-proton elastic scattering cross section at the kinematic point $(E_i, Q^2_i)$, evaluated with the 
DI$\chi$EFT FFs $G_E^p$ and $G_M^p$ with the parameters $(r_E^p, r_M^p)$ (the expression of the
elastic scattering cross section in terms of the FFs is given in Ref.~\cite{Bernauer:2013tpr}).
$\sigma_{{\rm exp}, i}$ is the measured cross section and $\Delta \sigma_{{\rm exp}, i}$
is the random uncertainty. The data points are grouped in $N_{\rm set}$ sets measured under the same 
running conditions; the normalization is assumed to be constant inside each set, but its value is unknown. 
The $N_{\rm set}$ parameters $(\Lambda_1, ..., \Lambda_{N_{\rm set}})$ represent the floating normalizations
in each set; $k(i)$ denotes the index $k$ of the set to which data point $i$ belongs. (A detailed 
discussion of how the experimental normalizations were defined and obtained can be found in 
Ref.~\cite{Higinbotham:2019jzd}.)
The $\chi^2$ defined by Eq.~(\ref{chi2}) is thus a function of the theory parameters $(r_E^p, r_M^p)$
and the normalization parameters $(\Lambda_1, ..., \Lambda_{N_{\rm set}})$. Minimization is performed
with respect to all the parameters simultaneously. The values of the $\Lambda_k (k = 1...N_{\rm set})$ at the 
minimum are found to be equal to unity within $\lesssim 1\%$; this indicates that the normalization determined 
in the original analysis of Ref.~\cite{Bernauer:2013tpr} is reproduced reasonably by our fit; 
the values themselves have no physical significance otherwise (nuisance parameters). 
The values of $(r_E^p, r_M^p)$ at the minimum correspond to the best fit to the data 
and represent the proton radii extracted with our method. 

%
% FIGURE
%
\begin{figure}[t]
\includegraphics[width=.9\columnwidth]{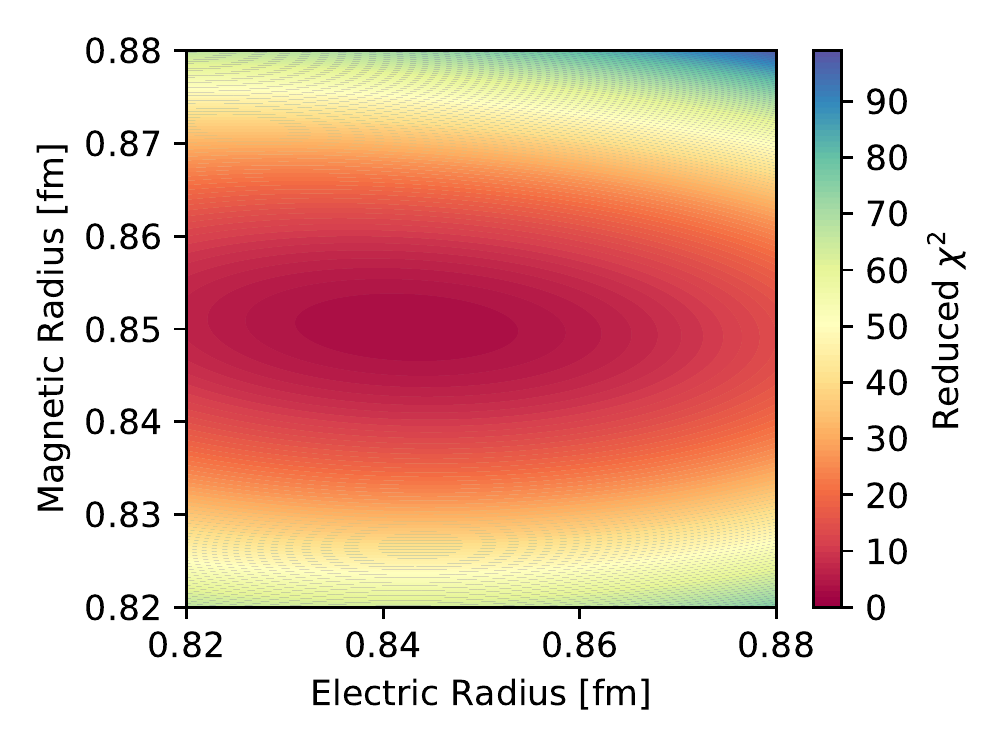}
\caption{Reduced $\chi^2$, Eq.~(\ref{chi2}), as a function of $r_E^p$ and $r_M^p$, after minimization 
with respect to the normalization parameters.}
\label{fig:chi2}
\end{figure}
To estimate the uncertainties of the extracted radii, we use the criterion $\Delta \chi^2$ = 2.7 to 
determine the 68\% confidence interval, corresponding to the simultaneous estimation of two independent parameters.
The uncertainties of the physical parameters $r_E^p$ and $r_M^p$ are affected also by the statistical 
fluctuations of the normalization parameters $\Lambda_k$; we have estimated the total statistical uncertainties 
using a bootstrap method and found them to be very close to the uncertainties of $r_E^p$ and $r_M^p$ 
one would obtain from the variation of the reduced $\chi^2$, 
obtained after minimization with respect to  $(\Lambda_1, ..., \Lambda_{N_{\rm set}})$.
In the final uncertainty we also include the theoretical uncertainty of the DI$\chi$EFT FFs 
(see below) \cite{Alarcon:2018zbz}.
\section{Results}
In the fit we include the cross section data up to a maximum momentum transfer, $Q^2 < Q^2_{\rm max}$.
Suitable values of $Q^2_{\rm max}$ are determined by considering the balance of experimental and 
theoretical uncertainties and the sensitivity of the cross sections to the model parameters
$r_E^p$ and $r_M^p$ \cite{Alarcon:2018zbz}.
Our standard fit uses $Q^2_{\rm max} = 0.5$ GeV$^2$ and includes 1285 of the 1422 Mainz A1 data points.
The overall quality of the description of the experimental cross sections is shown in
Fig.~\ref{fig:data} (the plots show also the data at $Q^2 > Q^2_{\rm max}$,
which were not included in the fit). One sees that all features of the kinematic dependence of the data (with the 
floating normalization determined by the fit) are reproduced by the theoretical model. The reduced
$\chi^2$ profile in the physical parameters $r_E^p$ and $r_M^p$, obtained after minimization with 
respect to the normalization parameters, is shown in Fig.~\ref{fig:chi2}. 
One observes that the variations of $\chi^2$ in $r_E^p$ and $r_M^p$ are
approximately independent, and that clear minima are obtained in both parameters.
Minimizing with respect to the radii, we extract $r_E^p = 0.842 \pm 0.002$ fm and 
$r_M^p = 0.850 \pm 0.001$ fm with a reduced $\chi^2$ of 1.39. 

To test the robustness of the results we have performed fits with different values of $Q^2_{\rm max}$
and found little effect on the extracted radii. In fact, using the entire Mainz A1 data set up to 
$Q^2_{\rm max} = 1$ GeV$^2$ gives $r_E^p = 0.843 \pm 0.002$ fm and $r_M^p = 0.850 \pm 0.001$ fm with a 
reduced $\chi^2$ of 1.43. This shows that the theoretical model (evaluated with the 
``best'' value of the radii) accurately describes the $Q^2$ dependence of the data over the 
entire range considered here.

As a further test we have performed fits to the rebinned version of the Mainz A1 data of Ref.~\cite{Lee:2015jqa},
where the original data sets are rescaled to a common normalization using empirical functional forms,
including the effects on the random uncertainties.
The fit with $Q^2_{\rm max} = 0.5$ GeV$^2$ uses 569 of the 658 rebinned data points and
gives radii $r_E^p = 0.840 \pm 0.002$ fm and $r_M^p = 0.849 \pm 0.001$ fm with a reduced 
$\chi^2$ of 1.07, in good agreement with our fit to the original data. Extending $Q^2_{\rm max}$
to include all the rebinned data we obtain $r_E^p = 0.841 \pm 0.003$ fm and $r_M^p = 0.849 \pm 0.001$ 
fm with a reduced $\chi^2$ of 1.10, showing similar stability the fit to the original data.
Overall, the tests show that the extracted radii are not sensitive to the choice of data
sets used in the fits.

The Jefferson Lab PRad experiment has reported a new measurement of the electron-proton elastic 
cross section down to $Q^2 \sim 10^{-4}$ GeV$^2$, significantly extending 
the reach of earlier measurements \cite{Xiong:2019umf}. We have performed a fit including the PRad data in addition
to the Mainz A1 data and found no change in the extracted $r_E^p$ and $r_M^p$ within uncertainties.
This happens because the DI$\chi$EFT model naturally describes the $Q^2$ dependence of the low-$Q^2$ data, 
with the same value of $r_E^p$ as favored by the higher-$Q^2$ data \cite{Alarcon:2018zbz};
this was also observed in the analysis of Ref.~\cite{Horbatsch:2019wdn}.
Note that the low-$Q^2$ data are sensitive mostly to $r_E^p$, and that our present extraction
of $r_M^p$ requires us to include data up to $Q^2 \sim 0.5$ GeV$^2$.

In our assessment of the errors of the extracted radii we must include also the intrinsic 
theoretical uncertainty of the DI$\chi$EFT model. This refers to the uncertainty in the predictions
for $G_E^p$ and $G_M^p$ for given values of $r_E^p$ and $r_M^p$, 
which results from the modeling of the high-mass states in the dispersion integral (effective poles) 
and was estimated in Ref.~\cite{Alarcon:2018zbz}. Performing fits with different values of the
effective pole mass we estimate the effect on the extracted radii as $\sim$ $\pm$0.010 fm for both 
$r_E^p$ and $r_M^p$; the interval should be regarded as the ``full plausible range'' of the
theoretical uncertainty. Our final results for the extracted radii are thus
$r_M^p$ = 0.850 $\pm$0.001 (fit 68\%) $\pm$0.010 (theory full range) fm 
and $r_E^p$ = 0.842 $\pm$0.002 (fit 68\%) $\pm$0.010 (theory full range) fm.
\section{Discussion}
Several aspects of our method and results merit further discussion. In our theory-based extraction
method the main impact on the radii comes from the data at ``higher'' $Q^2\sim$ 0.1--0.5 GeV$^2$
(see Fig.~\ref{fig:chi2} and Ref.~\cite{Alarcon:2018zbz}).
One observes that the magnetic radius is actually better determined than 
the electric one (see Fig.~\ref{fig:chi2}), because the cross section at the ``higher'' $Q^2$
is dominated by the contribution of the magnetic FF. In contrast, in methods based 
on the $Q^2 \rightarrow 0$ extrapolation the cross section is always dominated 
by the electric FF, rendering extraction of the magnetic radius extremely difficult.
Our method therefore offers principal advantages for the analysis of the proton's 
magnetic structure.

The results of our analysis validate previous results for the proton magnetic radius
$\sim$0.85 fm, obtained using dispersive fits of the earlier world data; 
see Ref.~\cite{Belushkin:2006qa} and references therein. They disagree with the 
results obtained from various empirical fits of the Mainz A1 data \cite{Bernauer:2010wm,Lee:2015jqa}.
This indicates that the observed discrepancies are due to the extraction methods
(analyticity, correlations between $Q^2$ regions from dispersion relations) 
rather than the different data sets.

The values of the electric and magnetic radii extracted from the data are very close.
While this may be accidental, it is qualitatively consistent with the nonrelativistic quark 
model picture (independent particle motion in an $L = 0$ orbital, no spin-orbit interactions).
Using our method we can also determine the proton's transverse charge and magnetization radii, 
which refer to the relativistic representation of the FFs in terms of transverse densities and can be related 
to the generalized parton distributions \cite{Miller:2018ybm,Diehl:2003ny,Belitsky:2005qn,Boffi:2007yc}.
The derivatives at $Q^2 = 0$ of the Dirac and Pauli FFs, $F_1^p$ and $F_2^p$,
are related to those of the electric and 
magnetic FFs by [$\kappa^p \equiv F_2^p(0) = \mu^p - 1$ is the anomalous magnetic moment, $m$ is the proton mass]
\begin{align}
\frac{dF_1^p}{dQ^2}(0) &= \frac{dG_E^p}{dQ^2}(0) + \frac{\kappa^p}{4 m^2}, 
\label{F1_der} \\
\frac{1}{\kappa^p} \frac{dF_2^p}{dQ^2}(0) &= \frac{1}{\kappa^p}\left[ \frac{dG_M^p}{dQ^2}(0) 
- \frac{dG_E^p}{dQ^2}(0) \right] - \frac{1}{4 m^2} .
\label{F2_der}
\end{align}
In the transverse density representation these derivatives determine the mean squared transverse radii of 
the distributions of charge and magnetization in the proton \cite{Miller:2018ybm},
\begin{align}
-4 \frac{dF_1^p}{dQ^2}(0) = \langle b^2 \rangle_1^p, \hspace{2em}
-\frac{4}{\kappa^p} \frac{dF_2^p}{dQ^2}(0) = \langle b^2 \rangle_2^p .
\label{b2}
\end{align}
Equations~(\ref{r2}) and (\ref{F1_der})--(\ref{b2}) linearly relate $\langle b^2 \rangle_1^p$ 
and $\langle b^2 \rangle_2^p$ to $(r_E^p)^2$ and $(r_M^p)^2$. Using the results of our fit, we obtain 
$\langle b^2 \rangle_1 = 0.394 \pm 0.002$ (fit 68\%) $\pm 0.011$ (theory full range) fm$^2$ 
and
$\langle b^2 \rangle_2 = 0.531 \pm 0.002$ (fit) $\pm 0.019$ (theory) fm$^2$. 
It is interesting to note that, if one neglected the small difference between the extracted
electric and magnetic radii and set $(r_E^p)^2 = (r_M^p)^2$, the transverse charge and magnetization
radii would be related as
\begin{align}
\langle b^2 \rangle_2^p - \langle b^2 \rangle_1^p = \mu^p/m^2
\hspace{2em} [\textrm{if $(r_E^p)^2 = (r_M^p)^2$}], 
\label{b2_1_2_difference}
\end{align}
i.e., the difference would be entirely proportional to the proton magnetic moment. 
Equation~(\ref{b2_1_2_difference}) is the partonic expression of the approximate equality 
of the electric and magnetic radii.
\begin{acknowledgments}
This material is based upon work supported by the U.S.~Department of Energy, 
Office of Science, Office of Nuclear Physics under contract DE-AC05-06OR23177.
J.M.A.\ acknowledges support from the Community of Madrid through the
Programa de atracci\'on de talento investigador 2017 (Modalidad 1),
the Spanish MECD grants FPA2016-77313-P.
\end{acknowledgments}
\bibliography{magnetic}
\end{document}